\documentclass[superscriptaddress,12pt]{revtex4}%
\usepackage{amssymb}
\usepackage{natbib}
\usepackage{anysize}
\usepackage{graphicx}
\usepackage{amsmath}
\usepackage[bookmarks = false]{hyperref}
\usepackage{amsfonts}
\usepackage{color}%
\marginsize{2.3cm}{2.3cm}{1.5cm}{1.5cm}

\begin{document}
\title{Evidence for association of triatomic molecule in ultracold $^{23}$Na$^{40}$K and $^{40}$K mixture}
\author{Huan Yang}
\thanks{These authors contributed equally to this work.}
\affiliation{Hefei National Laboratory for Physical Sciences at the Microscale and Department of Modern Physics, University of Science and Technology of China,
Hefei, Anhui 230026, China}
\affiliation{Shanghai Branch, CAS Center for Excellence and Synergetic Innovation Center in Quantum  Information
and Quantum Physics, University of Science and Technology of China, Shanghai 201315, China}
\affiliation{Shanghai Research Center for Quantum Sciences, Shanghai 201315, China}
\author{Xin-Yao Wang}
\thanks{These authors contributed equally to this work.}
\affiliation{Hefei National Laboratory for Physical Sciences at the Microscale and Department
of Modern Physics, University of Science and Technology of China,
Hefei, Anhui 230026, China}
\affiliation{Shanghai Branch, CAS Center for Excellence and Synergetic Innovation Center in Quantum  Information
and Quantum Physics, University of Science and Technology of China, Shanghai 201315, China}
\affiliation{Shanghai Research Center for Quantum Sciences, Shanghai 201315, China}
\affiliation{Beijing National Laboratory for Molecular Sciences, Key Laboratory of Molecular Nanostructure
and Nanotechnology, CAS Research/Education Center for Excellence in Molecular Sciences,
Institute of Chemistry, Chinese Academy of Sciences, Beijing 100190, China}
\affiliation{University of Chinese Academy of Sciences, Beijing 100049, China}
\author{Zhen Su}
\affiliation{Hefei National Laboratory for Physical Sciences at the Microscale and Department
of Modern Physics, University of Science and Technology of China,
Hefei, Anhui 230026, China}
\affiliation{Shanghai Branch, CAS Center for Excellence and Synergetic Innovation Center in Quantum  Information
and Quantum Physics, University of Science and Technology of China, Shanghai 201315, China}
\affiliation{Shanghai Research Center for Quantum Sciences, Shanghai 201315, China}
\author{Jin Cao}
\affiliation{Hefei National Laboratory for Physical Sciences at the Microscale and Department
of Modern Physics, University of Science and Technology of China,
Hefei, Anhui 230026, China}
\affiliation{Shanghai Branch, CAS Center for Excellence and Synergetic Innovation Center in Quantum  Information
and Quantum Physics, University of Science and Technology of China, Shanghai 201315, China}
\affiliation{Shanghai Research Center for Quantum Sciences, Shanghai 201315, China}
\author{De-Chao Zhang}
\affiliation{Hefei National Laboratory for Physical Sciences at the Microscale and Department
of Modern Physics, University of Science and Technology of China,
Hefei, Anhui 230026, China}
\affiliation{Shanghai Branch, CAS Center for Excellence and Synergetic Innovation Center in Quantum  Information
and Quantum Physics, University of Science and Technology of China, Shanghai 201315, China}
\affiliation{Shanghai Research Center for Quantum Sciences, Shanghai 201315, China}
\author{Jun Rui}
\affiliation{Hefei National Laboratory for Physical Sciences at the Microscale and Department
of Modern Physics, University of Science and Technology of China,
Hefei, Anhui 230026, China}
\affiliation{Shanghai Branch, CAS Center for Excellence and Synergetic Innovation Center in Quantum  Information
and Quantum Physics, University of Science and Technology of China, Shanghai 201315, China}
\affiliation{Shanghai Research Center for Quantum Sciences, Shanghai 201315, China}
\author{Bo Zhao}
\affiliation{Hefei National Laboratory for Physical Sciences at the Microscale and Department
of Modern Physics, University of Science and Technology of China,
Hefei, Anhui 230026, China}
\affiliation{Shanghai Branch, CAS Center for Excellence and Synergetic Innovation Center in Quantum  Information
and Quantum Physics, University of Science and Technology of China, Shanghai 201315, China}
\affiliation{Shanghai Research Center for Quantum Sciences, Shanghai 201315, China}
\author{Chun-Li Bai}
\affiliation{Beijing National Laboratory for Molecular Sciences, Key Laboratory of Molecular Nanostructure
and Nanotechnology, CAS Research/Education Center for Excellence in Molecular Sciences,
Institute of Chemistry, Chinese Academy of Sciences, Beijing 100190, China}
\affiliation{University of Chinese Academy of Sciences, Beijing 100049, China}
\author{Jian-Wei Pan}
\affiliation{Hefei National Laboratory for Physical Sciences at the Microscale and Department
of Modern Physics, University of Science and Technology of China,
Hefei, Anhui 230026, China}
\affiliation{Shanghai Branch, CAS Center for Excellence and Synergetic Innovation Center in Quantum  Information
and Quantum Physics, University of Science and Technology of China, Shanghai 201315, China}
\affiliation{Shanghai Research Center for Quantum Sciences, Shanghai 201315, China}

\maketitle

\textbf{Ultracold assembly of diatomic molecules has enabled great advances in controlled chemistry, ultracold chemical physics, and quantum simulation with molecules \cite{Krems2008,Carr2009, Quemener2012}. Extending the ultracold association to triatomic molecules will offer many new research opportunities and challenges in these fields. A possible approach is to form triatomic molecules in the ultracold atom and diatomic molecule mixture by employing the Feshbach resonance between them \cite{chin2010,Koehler2006}.
Although the ultracold atom-diatomic-molecule Feshbach resonances have been observed recently \cite{Yang2019,Wang2021}, utilizing these resonances to form triatomic molecules remains challenging.
Here we report on the evidence of the association of triatomic molecules near the Feshbach resonances between $^{23}$Na$^{40}$K molecules in the rovibrational ground state and $^{40}$K atoms. We apply a radio-frequency pulse to drive the free-bound transition and monitor the loss of $^{23}$Na$^{40}$K molecules. The association of triatomic molecules manifests itself as an additional loss feature in the radio-frequency spectra, which can be distinguished from the atomic loss feature.
The binding energy of triatomic molecule is estimated from the measurement.
Our work is helpful to understand the complex ultracold atom-molecule Feshbach resonance and may open up an avenue towards the preparation and control of ultracold triatomic molecules.}

\vspace{0.5cm}

Ultracold molecules have attracted great research interests due to potential applications in chemical physics, quantum information processing and precision measurements \cite{Krems2008,Carr2009, Quemener2012}. Recently, remarkable progress has been achieved in preparing and studying ultracold diatomic molecules. Direct laser cooling has been successfully applied to the molecules which have nearly closed cycling transitions \cite{Shuman2010, Barry2014}, and a temperature of a few microkelvin has been achieved \cite{Cheuk2018,Caldwell2019,Ding2020}. Various alkali-metal diatomic molecules have been created from ultracold atoms in bulk gases \cite{Ni2008,Molony2014,Takekoshi2014,Park2015,Guo2016,Rvachov2017,seesselberg2018,LiuL2019,Voges2020} or in optical tweezers \cite{LiuLR2019,He2020}. The studies of ultracold collisions involving molecules have enabled the demonstration of quantum state controlled chemical reaction \cite{Ospelkaus2010a,HuM2019}, the observation of atom-molecule Feshbach resonance \cite{Yang2019}, and  the collisional cooling of ultracold molecules \cite{Son2020}. By suppressing the ultracold reaction, quantum degenerate diatomic molecular gases have been produced in bulk gases \cite{DeMarco2019} and in two-dimensional layers \cite{Giacomo2020}.

Extending the study of ultracold molecules to triatomic molecules or even more complex polyatomic molecules will open up many new research frontiers in physics and chemistry. For example, the ultracold triatomic molecule represents a three-body system under the most stringent quantum constraint. Such a quantum-mechanical three-body problem is notoriously difficult and thus poses great challenges to the theory of few-body physics.
The molecular potential felt by the nuclei in the triatomic molecule
is usually described by an anisotropic three-body potential energy surface, which is extremely difficult to calculate with high accuracy, since it requires solving a many-electron Schr\"{o}dinger equation.
The ultracold triatomic molecule provides an ideal platform to probe the three-body potential energy surface since the molecular level of the ultracold molecule can be measured with exceptional resolution.
Moreover, ultracold triatomic molecules have more symmetric properties and more degrees of freedom that can be controlled by external fields. Precise control of these degrees of freedom offers new opportunities in tests of fundamental physics and quantum simulation of exotic Hamiltonian.
Recently, laser cooling and optoelectrical cooling have been employed to cool the linear triatomic molecule \cite{Ivan2017,Baum2020} and other polyatomic molecules \cite{Zeppenfeld2012, Alexander2016,Debayan2020}, and the temperatures have been reduced to  the submillikelvin regime.



Besides direct cooling, another possible solution to prepare ultracold triatomic molecules is to associate them from the ultracold atom and diatomic molecule mixture. The major difficulty of ultracold association is that the coupling strength between the triatomic bound state and the atom-diatomic-molecule scattering state is usually very weak. This problem may be solved by employing the atom-diatomic-molecule Feshbach resonance, which occurs as the energy of the triatomic bound state coincides with the energy of the scattering state \cite{Yang2019,Qiu2006,Vogels2015,Chefdeville2013,Klein2017}. Therefore, close to the resonance, the coupling strength between the triatomic bound state and the atom-diatomic-molecule scattering state is resonantly enhanced. Moreover, at ultracold temperatures, Feshbach resonances can be tuned by external electromagnetic fields \cite{chin2010,Koehler2006}. This kind of controllability provides an essential tool to convert the atom and diatomic molecule pairs into the triatomic molecules close to the atom-diatomic-molecule Feshbach resonance. This technique has been widely used in the association of weakly bound diatomic molecules \cite{Regal2003,Herbig2003,Thompson2005,Ospelkaus2006} and the association of universal Efimov trimers \cite{Lompe2010b,Nakajima2011,Machtey2012}.

Here we explore the association of triatomic molecules in the mixture of ultracold ground-state $^{23}$Na$^{40}$K molecules and $^{40}$K  atoms. Although a number of Feshbach resonances between  $^{23}$Na$^{40}$K and $^{40}$K have been observed \cite{Yang2019,Wang2021}, understanding these resonances remains challenging, owing to the complexity of the ultracold atom-diatomic-molecule collision. A detailed experimental and theoretical study of the resonance pattern suggests that the triatomic bound state may be assigned as the long-range bound states that have similar character to the incoming scattering states \cite{Wang2021}. However, little is known about the free-bound coupling mechanism and the coupling strength. The lifetime of the triatomic molecule may be short, due to the collisions with the atoms and diatomic molecules, and the possible excitations by the trapping lasers \cite{Christianen2019}. In this case, we employ radio-frequency (rf) fields to associate the triatomic molecules. We apply an rf pulse to drive the free-bound transition in the vicinity of the Feshbach resonance between $^{23}$Na$^{40}$K molecules and $^{40}$K atoms, and monitor the loss of $^{23}$Na$^{40}$K molecules. The advantage of this method is that even if the lifetime of the triatomic molecule is short, the signature of association may still be observed from the loss of $^{23}$Na$^{40}$K molecules.
We present clear evidence of the formation of triatomic molecules by observing additional loss features distinct from atomic loss features in the rf spectra.
The binding energy of the triatomic molecule is estimated from the measurement.

Our experiment starts from a mixture of $^{23}$Na$^{40}$K molecules in the rovibrational ground state and $^{40}$K atoms confined in a crossed-beam optical dipole trap. The experimental setup has been described previously \cite{LiuL2019,Yang2019}. We first prepare $^{23}$Na and $^{40}$K atomic mixture at a temperature of about 600 nK, and create weakly bound $^{23}$Na$^{40}$K Feshbach molecules  near the atomic Feshbach resonance between $^{23}$Na and $^{40}$K. The $^{23}$Na$^{40}$K molecules are transferred to the rovibrational ground state by stimulated Raman adiabatic passage (STIRAP) and the $^{23}$Na atoms are then removed by resonant light pulses (see Methods). By selecting an appropriate hyperfine excited state and a proper laser polarization, we can prepare $^{23}$Na$^{40}$K molecules in a specific hyperfine state labelled by $|v=0,n=0,m_{\rm{Na}},m_{\rm{K}}\rangle$, where $v$ and $n$ are the vibrational and rotational quantum numbers, and $m_{\rm{Na}}$ and  $m{_{\rm{K}}}$ represent the projections of the nuclear spins along the direction of the magnetic field. The $^{40}$K atoms are transferred to the $|9/2,-9/2\rangle$ state. As illustrated in Fig. 1, we utilize the Feshbach resonance between the $|0,0,m_{\rm{Na}},m_{\rm{K}}\rangle$ and $|9/2,-7/2\rangle$ states to form the triatomic molecules, by driving the atomic transition $|9/2,-9/2\rangle\rightarrow|9/2,-7/2\rangle$ through rf fields.

In the vicinity of the magnetic Feshbach resonance between $^{23}$Na$^{40}$K  and $^{40}$K, if the frequency of the rf field is on resonance with the bare atomic transition, the $^{40}$K atoms are transferred to the $|9/2,-7/2\rangle$ state. The $^{23}$Na$^{40}$K molecules will be quickly lost due to the inelastic collisions with the $|9/2,-7/2\rangle$ atoms, which is resonantly enhanced near the Feshbach resonance. This occurs no matter whether the magnetic field is above or below the resonance position. However, on the molecular side of the resonance, the triatomic bound state exists since its energy is below the scattering threshold. 
At these magnetic fields, if the frequency of the rf field is on resonance with the free-bound transition, the triatomic molecules may be formed in the $^{23}$Na$^{40}$K and $^{40}$K mixture. The triatomic molecules are in an unstable vibrational excited state. They may decay quickly due to the collisions with the atoms and molecules or the excitations by the trapping lasers. Consequently, the association of the triatomic molecule will also lead to a loss of $^{23}$Na$^{40}$K molecules.
The association of the triatomic molecules will manifest itself as an additional loss feature on the rf loss spectrum, whose frequency is below the atomic loss feature.

We prepare $^{23}$Na$^{40}$K molecules in the $|0,0,-3/2,-3\rangle$ state at about 84.9 G. There is a Feshbach resonance between the $|0,0,-3/2,-3\rangle$ and $|9/2,-7/2\rangle$  states located at about 57.6 G with a width of about 5.3 G \cite{Wang2021}. According to the analysis in Ref. \cite{Wang2021}, we assume the triatomic molecule exists at the magnetic fields below the resonance position. After the $^{23}$Na$^{40}$K and $^{40}$K mixture is prepared, we ramp the magnetic field to a target value $B_t$ close to the Feshbach resonance in 3 ms and wait 15 ms for the magnetic field to stabilize. We then apply a 30-ms rf pulse with a frequency close to the atomic transition $|9/2,-9/2\rangle\rightarrow|9/2,-7/2\rangle$ to associate the triatomic molecules. The Rabi frequency of the rf pulse is about $2\pi\times20-30$ kHz.  After that, the $^{40}$K atoms are removed by resonant light pulses. We ramp the magnetic field back to the initial value and transfer the remaining $^{23}$Na$^{40}$K molecules to the Feshbach state by a reverse STIRAP for detection.

The remaining number of  $^{23}$Na$^{40}$K molecules as a function of the rf frequency are shown in Fig. 2 for different magnetic fields. The rf spectra are displayed relative to the atomic transition at $B_t$. At the magnetic fields $57.114\leq B_t\leq58.091$ G, the loss due to the atomic transfer is observed. The atomic loss feature can be well fitted to a Gaussian function with the center being a few kHz from the atomic transition. This offset is at least one order of magnitude smaller than the width of the atomic loss feature and is mainly caused by the drifts and fluctuations of the magnetic field. The observation of only the atomic loss feature at these magnetic fields is consistent with our assumption that these magnetic fields are either above the resonance position or very close to the resonance position if it is on the molecular side. However, at the magnetic fields $B_t\leq56.466$ G, an additional loss feature emerges on the left shoulder of the atomic loss feature. Due to the presence of this additional loss feature, the overall loss feature becomes asymmetric and broad. This additional loss feature can be clearly observed at magnetic fields up to $B_t=55.887$ G. At $B_t=55.756$ G, the signature of the the additional loss feature is weak and cannot be unambiguously resolved. We have checked that the additional loss features cannot be explained by the change of the magnetic field. During the application of rf pulses, the uncertainty of the magnetic field is less than about 30 mG (see Methods) and thus the change of the magnetic field will induce an uncertainty of the atomic transition by a few kHz. This loss feature cannot be caused by the mean field shift either, since the density of $^{23}$Na$^{40}$K molecules is about one order of magnitude smaller than the density of $^{40}$K atoms, and thus the mean field shift is negligible in our experiment. Based on these arguments and our understanding of the atom-diatomic-molecule Feshbach resonance, we interpret this additional loss feature as the evidence of the association of the triatomic molecules.

We perform similar experiments using the molecule hyperfine state $|0,0,-1/2,-4\rangle$, which has the same total projection $m_{\rm{Na}}+m_{\rm{K}}=-9/2$ as the $|0,0,-3/2,-3\rangle$ state. A Feshbach resonance between $|0,0,-1/2,-4\rangle$ and $|9/2,-7/2\rangle$  is located at about 57.5 G with a width of about 2.5 G \cite{Wang2021}. It is very close to the resonance between $|0,0,-3/2,-3\rangle$ and $|9/2,-7/2\rangle$ studied above. Therefore, similar experimental methods are applied. The measured rf spectra are shown in Fig. 3. At $B_t=57.593$ and 57.114 G, only the atomic loss feature is observed. However, the association feature can be clearly observed at the magnetic fields $55.821\leq B_t\leq 56.121$ G.

The measurement of the rf spectra may allow us to estimate the binding energy of the triatomic molecule. Since there is no available model to describe the triatomic bound state, we simply use two overlapping Gaussian functions to fit the data, where the Gaussian function for the atomic loss feature is centered at the atomic transition at $B_t$. The binding energy is given by the central position of the Gaussian function for the association feature, and is determined by the fit. The binding energies determined in this way are shown in Fig. 4. For the molecular state $|0,0,-3/2,-3\rangle$, the binding energies between $B_t=56.466$ G and 55.756 G  are given. For the molecular state $|0,0,-1/2,-4\rangle$, the binding energies between $B_t=56.121$ G and 55.821 G are given.   As a reference, the corresponding resonance positions obtained from the loss measurements are also shown. 
We find the binding energy determined from the fit is not accurate enough to see how the association feature changes with the magnetic field. This is mainly because we have used a strong and long rf pulse to make the association features clearly visible. Consequently, the atomic loss features and association features are broadened and overlapped with each other. In this case, the binding energy determined from the fit is sensitive to the central position of the atomic loss feature, which is fixed at the atomic transition in our model. Besides, other systematic uncertainties, such as the fitting model, power broadening and finite temperature effects also affect the accuracy of the binding energy.  Therefore, the binding energy given here is a rough estimate. The accuracy of the binding energy may be further improved by increasing the signal-to-noise ratio.

In conclusion, we have associated triatomic molecules in the ultracold $^{23}$Na$^{40}$K and $^{40}$K mixture using rf fields. The rf association provides a spectroscopic probe of the triatomic molecule close to the Feshbach resonance between $^{23}$Na$^{40}$K and $^{40}$K. These observations may help to understand the complex ultracold atom-molecule Feshbach resonance, which is much more challenging to understand than the atomic Feshbach resonances.
Our work makes an important step towards preparing an ensemble of ultracold triatomic molecules. Achieving this goal requires  the lifetime of triatomic molecules in the optical dipole trap is long enough so that we can isolate them from $^{23}$Na$^{40}$K molecules and $^{40}$K atoms. The triatomic molecules formed close to the Feshbach resonance are weakly bound molecules near the atom-diatomic-molecule scattering threshold, and they may be transferred to deeply bound states by lasers.

\section*{Methods}

\noindent
\textbf{Prepation of $^{23}$Na$^{40}$K  molecules}

\noindent
Our experiment starts with approximately $2.8\times10^5$ $^{23}$Na and $1.0\times10^5$ $^{40}$K atoms confined in a crossed-beam optical dipole trap at a temperature of about 600 nK.
The ultracold atomic mixture is prepared as follows. We load two-species dark spot MOT by employing a $^{23}$Na Zeeman slower and a $^{40}$K 2D MOT. The atoms are transferred to a magnetic trap to perform evaporative cooling. The atoms are then loaded into a crossed-beam optical dipole trap with a wavelength of 1064 nm, and are further  cooled by reducing the power of the trapping lasers. The trap frequencies for $^{40}$K atoms are $2 \pi\times (250,239,72)$ Hz.

After the ultracold atomic mixture is created, we employ the atomic Feshbach resonance between the $|f,m_{f}\rangle_{\rm{Na}}=|1,1\rangle$ and $|f,m_{f}\rangle_{\rm{K}}=|9/2,-9/2\rangle$ states at about 89 G \cite{Park2012} to create $^{23}$Na$^{40}$K Feshbach molecules. We prepare the atoms in the $|1,1\rangle$ and $|9/2,-7/2\rangle$ states and create $^{23}$Na$^{40}$K  Feshbach molecules at about 84.9 G using two-photon Raman photoassociation \cite{Yang2019}. The Raman light couples the $|9/2,-7/2\rangle\rightarrow|9/2,-9/2\rangle$ transition with a Rabi frequency of about $2\pi\times$45 kHz. We typically produce about $1.2\times 10^4$ $^{23}$Na$^{40}$K  Feshbach molecules. The $^{23}$Na$^{40}$K Feshbach molecules are then transferred to the rovibrational ground state by stimulate Raman adiabatic passage (STIRAP). We employ the mixture of $B^{1}\Pi$ $|v=12,J=1\rangle$ and $c^{3}\Sigma$ $|v=35,J=1\rangle$ molecule
electronic excited states \cite{Park2015} as the intermediate states. The
pump laser (805 nm) that couples the Feshbach state and the intermediate state, and the Stokes laser (567 nm) that couples the intermediate state and the rovibrational ground state are locked to an ultralow expansion  cavity to suppress the noises. We use the hyperfine level of the excited state $F_1=3/2, m_{F_1}=-3/2, m_{\rm{K}}=-3$ as the intermediate state to prepare $^{23}$Na$^{40}$K molecules in the hyperfine level $|0,0,-3/2,-3\rangle$ of the ground state. The hyperfine level $|0,0,-1/2,-4\rangle$ is prepared by using the hyperfine level of the excited state $F_1=1/2, m_{F_1}=-1/2, m_{\rm{K}}=-4$ as the intermediate state.  In both cases, the $\pi$ polarized Stokes laser propagates perpendicular to the direction of the magnetic field,  and the $\sigma^{-}$ polarized pump laser propagates along the direction of the magnetic field.  The Rabi frequencies for Stokes and pump light are about 2-3 MHz and the efficiency of a round-trip STIRAP is about 40-50\%.
After  the ground-state $^{23}$Na$^{40}$K molecules are created, we remove the remaining $^{23}$Na atoms by resonant light pulses, and transfer the $^{40}$K atoms to the $|9/2,-9/2\rangle$ state. In this way, the ultracold $^{23}$Na$^{40}$K and $^{40}$K  mixture is prepared.

\bigskip

\noindent
\textbf{Magnetic field}

\noindent
After the preparation of the $^{23}$Na$^{40}$K and $^{40}$K  mixture, we ramp the magnetic field to a target value $B_t$ in 3 ms and then wait 15 ms for the magnetic field to stabilize. We have programmed the profile of the current of the magnetic coils in order to compensate the eddy currents induced by the stainless steel chamber and the coils. The $^{23}$Na$^{40}$K and $^{40}$K mixture is held at the target magnetic field and we apply a 30-ms rf pulse to associate the triatomic molecules. The $^{40}$K atoms are then removed by resonant light pulses and we ramp the magnetic field back to the initial value in 3 ms. We wait about 26 ms for the magnetic field to stabilize and then transfer the ground-state $^{23}$Na$^{40}$K molecules to the Feshbach state and detect the remaining $^{23}$Na$^{40}$K molecules by standard absorption imaging. The evidence for association of the triatomic molecules have been shown in Figs. 2 and 3 of the main text.


To make sure the association feature is not caused by the change of the magnetic field, we measure the magnetic field as a function of the hold time by performing rf spectroscopy on the $|9/2,-5/2\rangle\rightarrow|9/2,-7/2\rangle$ transition. To include the effects of the association rf pulse on the magnetic field, we use the same time sequence as the association experiment, except that the loading time of $^{23}$Na atoms is short so that after the evaporative cooling only $^{40}$K atoms remain in the optical trap. The frequency of the association pulse is  chosen to be about 100 kHz lower than the free atomic transition $|9/2,-7/2\rangle\rightarrow|9/2,-9/2\rangle$, so that it will not affect the rf spectroscopy on the $|9/2,-5/2\rangle\rightarrow|9/2,-7/2\rangle$ transition. The magnetic field at the various hold time is measured immediately after the truncation of the association pulse. The magnetic field as a function of the hold time is shown in Extended Data Figure 1. We find that the magnetic field drifts and fluctuates during the hold time, and the uncertainty is less than about 20 mG from the target magnetic field. Besides this uncertainty, there is also a long-term fluctuation of the magnetic field, which is estimated to be about 10 mG. Therefore, we estimate the overall uncertainty of the magnetic field is about 30 mG.

\renewcommand\subsection[1]{
\vspace{\baselineskip}
\textbf{#1}
\vspace{0.5\baselineskip}}

\subsection{Acknowledgement}

This work was supported by the National  Key R\&D Program of China (under Grant No.
2018YFA0306502), the National Natural Science Foundation of China (under Grant No. 11521063, 11904355),  the
Chinese Academy of Sciences, the Anhui Initiative in Quantum Information Technologies, Shanghai Municipal Science and Technology Major Project (Grant No.2019SHZDZX01), Shanghai Rising-Star Program (Grant No. 20QA1410000).

\newpage

\begin{figure}[tbh]
\centering
\includegraphics[width=10cm]{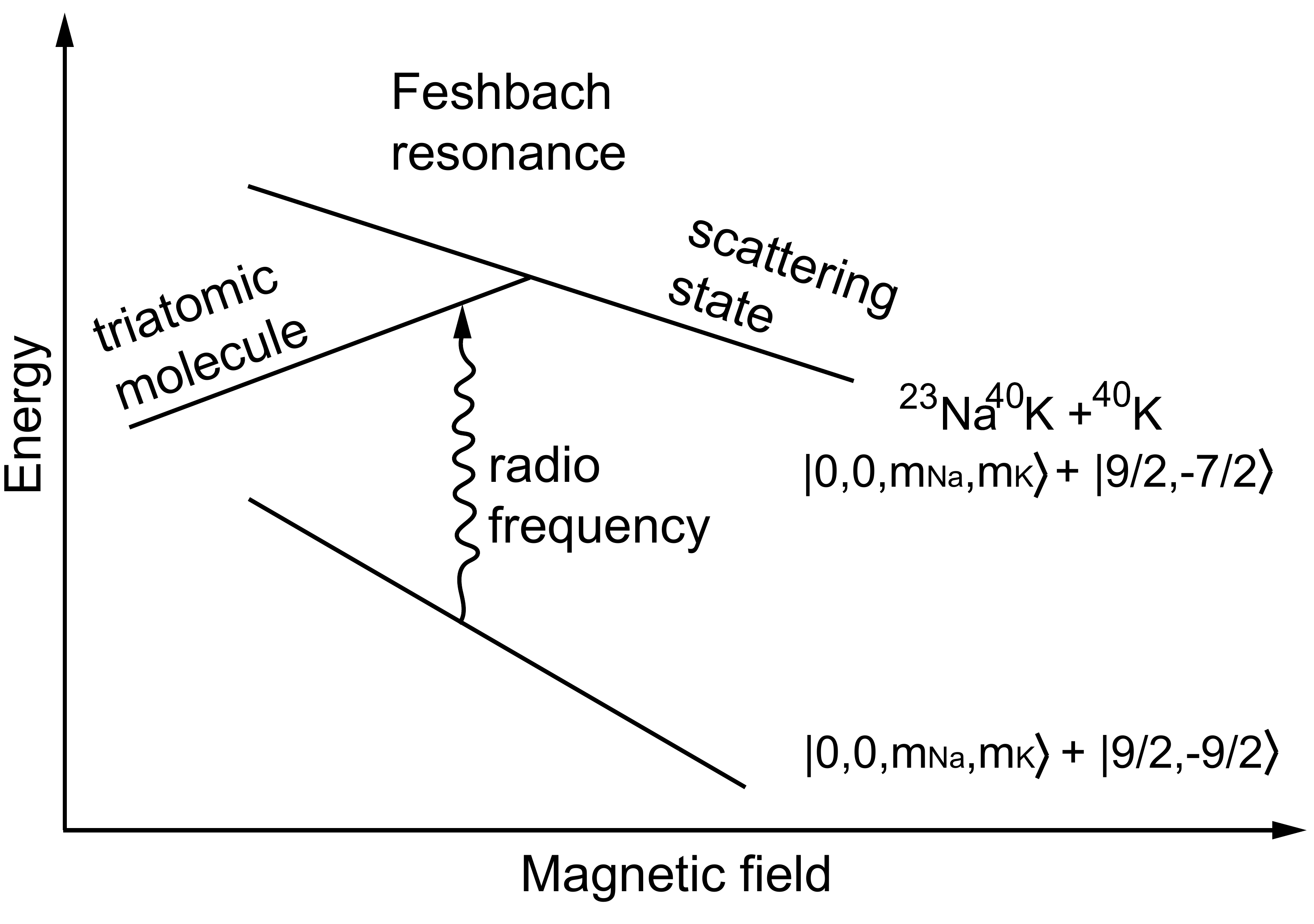}
\caption{\textbf{Illustration of radio-frequency association of triatomic molecules in the vicinity of the Feshbach resonance between $^{23}$Na$^{40}$K and $^{40}$K}. The Feshbach resonance  occurs at the magnetic field where the energy of the triatomic molecule coincides with the energy of the scattering state $|0,0,m_{\rm{Na}},m_{\rm{K}}\rangle$+$|9/2,-7/2\rangle$. The triatomic molecule exists at the magnetic fields where its energy is smaller than the energy of the  scattering state. Starting from a different scattering state $|0,0,m_{\rm{Na}},m_{\rm{K}}\rangle$+$|9/2,-9/2\rangle$, the triatomic molecule may be formed by applying a radio-frequency field on resonance with the free-bound transition.}%
\label{fig3}%
\end{figure}

\begin{figure}[tbh]
\centering
\includegraphics[width=12cm]{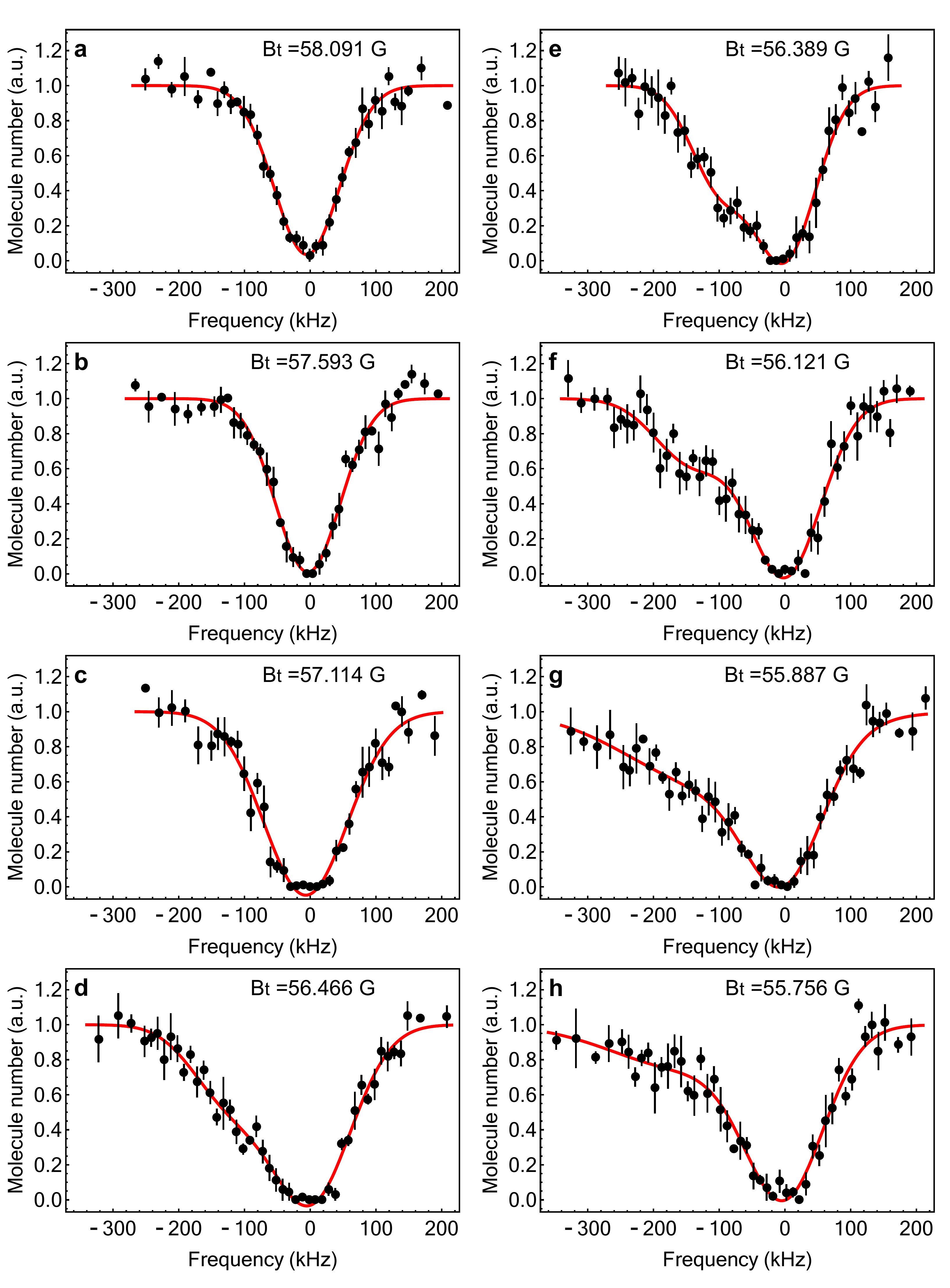}
\caption{\textbf{Radio-frequency spectra measured at different magnetic fields near the Feshbach resonance between $|0,0,-3/2,-3\rangle$ and $|9/2,-7/2\rangle$}. The remaining number of $^{23}$Na$^{40}$K molecules are plotted as a function of the rf frequency. The rf frequency is referenced to the atomic transition $|9/2,-9/2\rangle\rightarrow|9/2,-7/2\rangle$ at $B_t$. \textbf{a}-\textbf{c}, At these magnetic fields, the loss due to the atomic transfer is observed. The solid lines are Gaussian fits to the data. \textbf{d}-\textbf{h}, At these magnetic fields, an additional loss feature due to the formation of triatomic molecules appears on the left shoulder of the atomic loss feature. We use two overlapping Gaussian functions to fit the data, where the Gaussian function for the atomic loss feature is centered at the atomic transition. Each point represents the average of three to six measurements and error bars represent standard error of mean.}%
\label{fig3}%
\end{figure}

\begin{figure}[tbh]
\centering
\includegraphics[width=12cm]{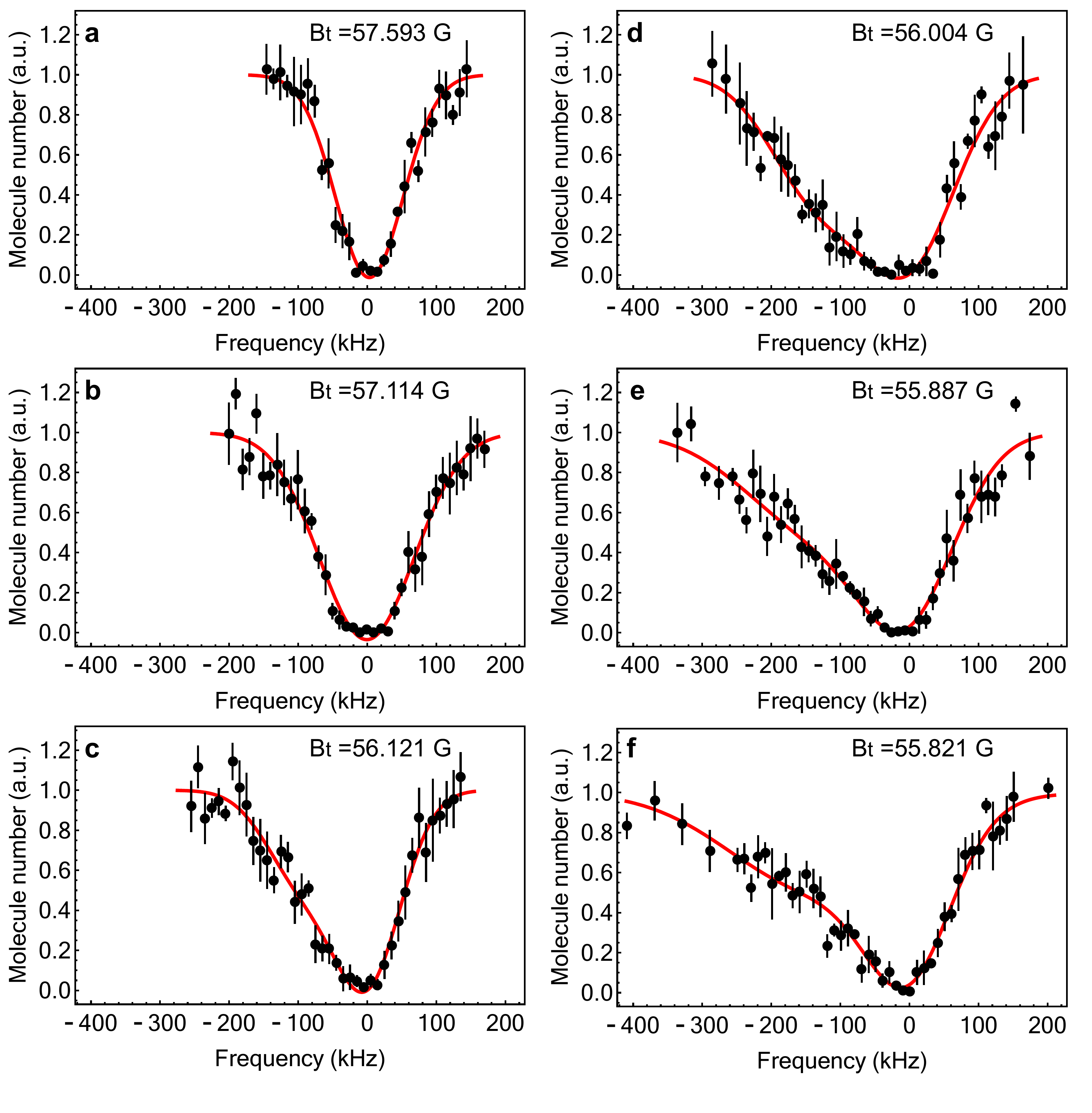}
\caption{\textbf{Radio-frequency spectra measured at different magnetic fields close to the Feshbach resonance between $|0,0,-1/2,-4\rangle$ and $|9/2,-7/2\rangle$}. The rf frequency is referenced to the atomic transition $|9/2,-9/2\rangle\rightarrow|9/2,-7/2\rangle$ at $B_t$. \textbf{a} and \textbf{b}, At these magnetic fields, the atomic loss feature is observed. The data points are fitted to a Gaussian function. \textbf{c}-\textbf{f}, At these magnetic fields, the association feature is clearly observed. The solid lines represent a fit of two overlapping Gaussian functions to the data, where the Gaussian function for the atomic loss feature is centered at the atomic transition. Each point represents the average of three to six measurements and error bars represent standard error of mean.}%
\label{fig3}%
\end{figure}

\begin{figure}[tbh]
\centering
\includegraphics[width=8cm]{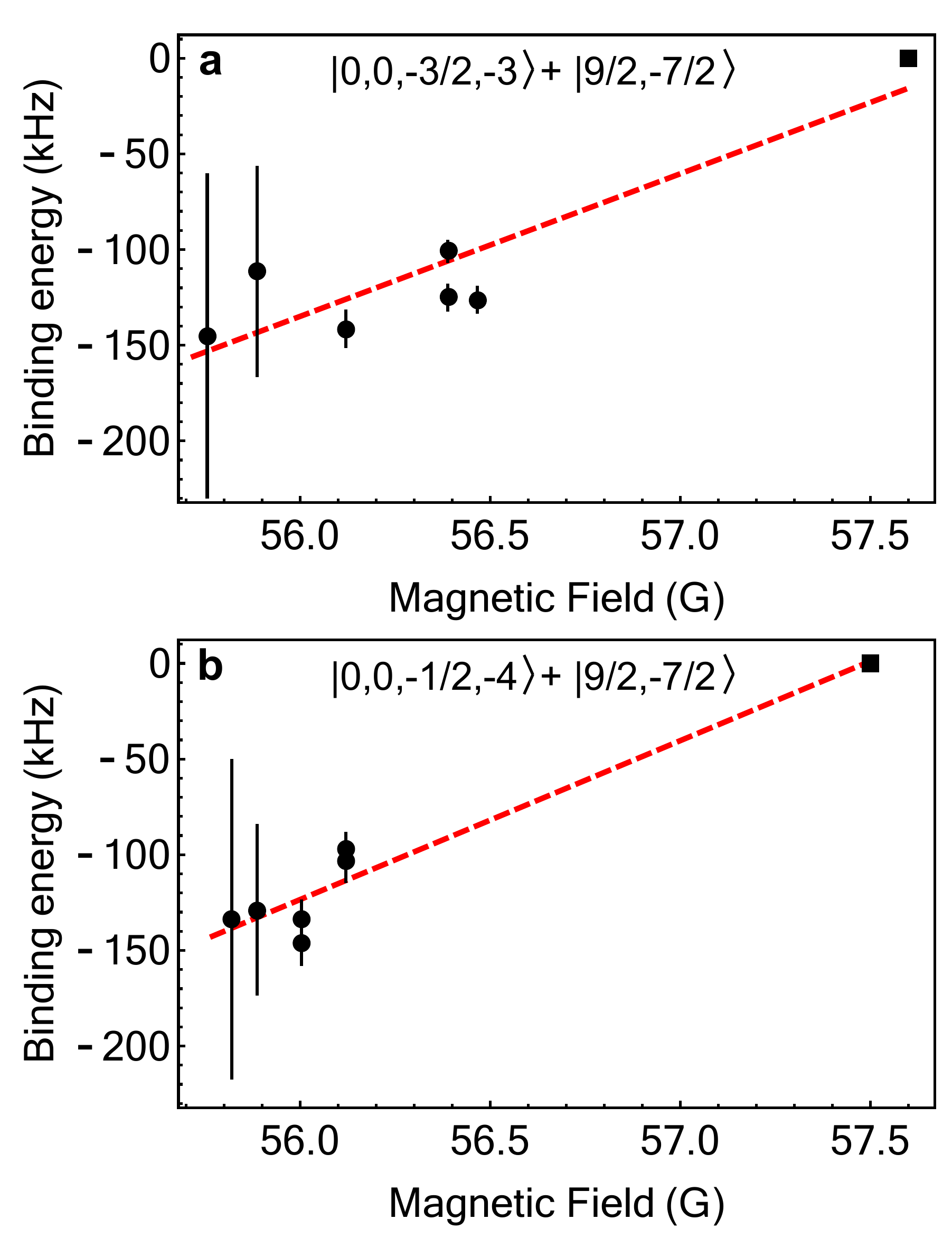}
\caption{\textbf{Binding energies of the triatomic molecules determined from the fit are plotted for different magnetic fields}. The binding energies of the triatomic molecules close to the Fesbach resonance in the collision channel \textbf{a}, $|0,0,-3/2,-3\rangle+|9/2,-7/2\rangle$ and \textbf{b}, $|0,0,-1/2,-4\rangle+|9/2,-7/2\rangle$. Error bars represent standard error of the fit. The resonance positions obtained from the loss measurements are marked by black squares. The dashed lines are introduced to guide the eye. }%
\label{fig3}%
\end{figure}

\newpage

\begin{figure}[pth]
\centering
\renewcommand\thefigure{Exteneded Data Figure 1}
\includegraphics[width=12cm]{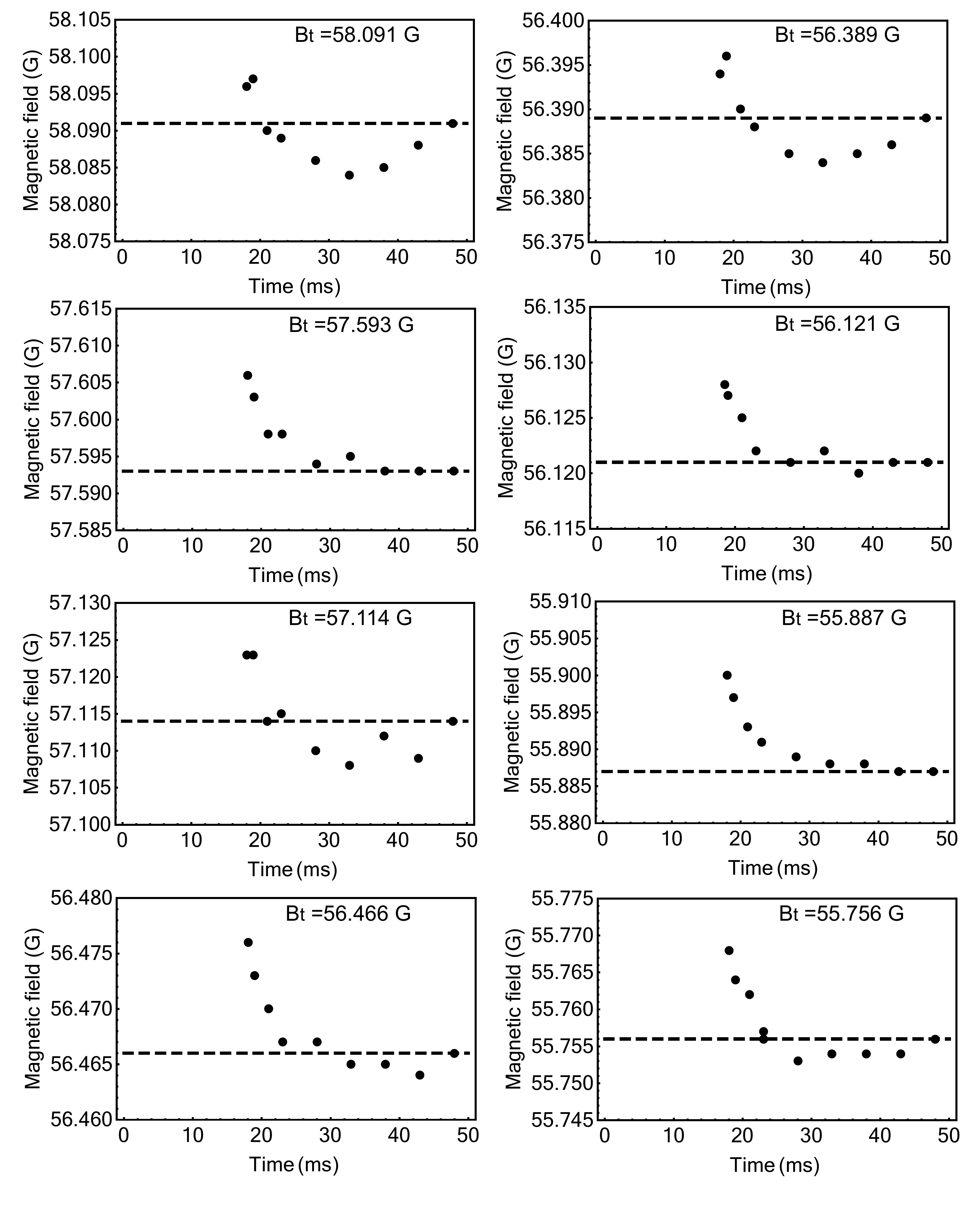}
\caption{ The magnetic field as a function of the hold time. The ground-state $^{23}$Na$^{40}$K molecules are created at $t=0$. The magnetic field is ramped to the target value $B_t$ in 3 ms and we wait 15 ms for the magnetic field to stabilize. The association rf pulse is applied at $t=18$ ms with a duration of 30 ms. The magnetic field between 18 ms and 50 ms is measured by rf spectroscopy. The dashed lines represent the target magnetic field $B_t$.}%
\label{figs3}%
\end{figure}

\end{document}